\documentclass[12pt,preprint]{aastex}
\shortauthors{del R\'{\i}o, Brinks \& Cepa}    
\shorttitle{\ion{H}{1} distribution in NGC~404}

\def\plotfiddle#1#2#3#4#5#6#7{\centering \leavevmode
\vbox to#2{\rule{0pt}{#2}}
\includegraphics{#1}}

\newcommand{\mhi}{M_{\rm H{\mbox{\scriptsize\sc i}}}}
\begin{document}
\title{High Resolution \ion{H}{1} Observations of the Galaxy NGC\,404:
a dwarf S0 with abundant Interstellar Gas}
\author{M. S. del R\'{\i}o}
\affil{Instituto Nacional de Astrof\'{\i}sica, Optica y Electr\'onica,
   Tonantzintla, 72000 Puebla, Mexico and Universidad Popular Aut\'onoma del 
   Estado de Puebla, 72160 Puebla, Mexico}
\email{sole@inaoep.mx}
\and
\author{E. Brinks}
\affil{Instituto Nacional de Astrof\'{\i}sica, Optica y Electr\'onica,
  Tonantzintla, 72000 Puebla, Mexico}
\email{ebrinks@inaoep.mx}
\and
\author{J. Cepa\altaffilmark{1}}
\affil{Instituto de Astrof\'{\i}sica de Canarias, E--38200 La Laguna, Tenerife,
  Spain}
\email{jcn@ll.iac.es}
\altaffiltext{1}{Departamento de Astrof\'{\i}sica, Facultad de F\'{\i}sica,
  Universidad de La Laguna, E--38071  La Laguna, Tenerife, Spain}

\begin{abstract}
As part of a detailed study of the gas content in a sample of Early-Type
galaxies, we present 21\,cm \ion{H}{1} line maps of the S0 galaxy
NGC\,404, at a spatial resolution of $15.2''\times14.4''
\, (\alpha\times \delta)$ and a velocity resolution of $\rm 2.6\,
km\,s^{-1}$. The \ion{H}{1} has been traced out to a radius R $\sim 8R_{25}$ or 
48 disc scale-lengths, making it one of the largest \ion{H}{1} extents
reported ($800''$ or 12.6\,kpc at the assumed distance of
3.3\,Mpc). Approximately 75\% of the \ion{H}{1} resides in a doughnut which
is seen close to face--on with inner and outer radii of $\sim R_{25}$ and
$\sim 4R_{25}$. The optical galaxy fits snugly within the hole of the
doughnut. The remaining 25\% of the neutral gas is found in an annulus concentric
with the doughnut and with a somewhat larger ellipticity, extending from a
radius of $\sim 4R_{25}$ to $\sim 8R_{25}$.  A total \ion{H}{1} mass of
$1.52\pm 0.04\times10^8\, M_\odot$ is found, which implies an $\mhi/L_{B} =
0.22$ in solar units. We argue that most if not all of this gas is of
external origin, most likely due to the merger of a dwarf irregular galaxy
with $M_B \sim -15.5$\,mag.

The velocity field shows a steeply declining observed rotation curve, 
compatible with Keplerian decline. However, because the galaxy is close to face--on
there is a degeneracy in the determination of the intrinsic rotation curve and
inclination. We therefore analysed two extreme cases, producing tilted--ring model 
fits forcing either a Keplerian or a Flat rotation curve through the observations;
both approaches result in equally plausible fits. In both model fits, the position angle 
of the kinematical
major axis of the annulus is distinct from that of the doughnut and ranges
from $160^\circ$ to $120^\circ$ (for the doughnut these values are
$100^\circ$ to $60^\circ$).

Assuming a distance of 3.3\,Mpc, a total mass of $3\times10^{10}\, M_\odot$
is found on the basis of the Keplerian rotation curve. This implies a global
$M_{\rm T}/L_{B}$ ratio of $\sim 44$ in solar units which is 
high and likely a reflection of the low blue luminosity of the galaxy ($\sim
15$ times lower than the average S0 luminosity). Values for a Flat rotation
curve are a factor of 4 higher.

\end{abstract}


\section{Introduction}

Traditionally, S0 galaxies were thought to represent a transition between
elliptical (E) and spiral (S) galaxies. They share with the latter a central
bulge and an exponential disc, and with the former the lack of spiral
structure or arms. They are considered inert objects with very little gas,
composed almost exclusively of Population II type stars, and virtually a
total absence of on--going star formation. This picture has changed
considerably over the past 20 years thanks mainly to tremendous progress in
observational techniques which are now routinely delivering high sensitivity
data over a wide range of wavelengths.

The interstellar medium of early--type galaxies, and especially that of S0s,
has attracted a great deal of interest ever since \citet{fg76} showed that a
considerable amount of mass ejected by evolved stars could pile up over the
lifetime of the galaxy. Surprisingly, though, high resolution, sensitive
\ion{H}{1} observations demonstrated that in general hardly any gas is
detected and that most of it seems to have been accreted
(\citealt{vgea86,wh90}). The scarcity of material detected in \ion{H}{1},
H$_\alpha$, CO, and at FIR wavelengths and the origin of that material form
two puzzles of contemporary astronomy.

\citet{fg76} discuss several ways in which gas could have been expelled or
eliminated from early--type systems. They reach the conclusion that gas is
particularly hard to remove from these systems via star formation. Although
star formation has virtually ceased, there are indications that some star
formation is taking place in S0s, as based on IRAS observations
(\citealt{j86,kgkj89}) and CO observations (\citealt{wh89,s90}) which show
that some S0s rival spirals in their CO content, but that the molecular gas
is concentrated within 1--2 kpc from the centre.

We have started a project, mapping a few early--type galaxies in detail, in
order to shed light on the origin of the ISM in these objects. In this paper
we present the results obtained for one of the galaxies under study,
NGC\,404.

NGC\,404 is a typical example of an early--type galaxy which shows more
activity than expected from such systems. It contains an appreciable amount
of atomic \citep{bw76} and molecular gas \citep{wh90}, a prominent,
semicircular dust lane encircling the nucleus \citep*{bcr82} and various
bright UV point sources \citep{mea98}. It was classified a LINER by
\citet{sea90}. \citet{m58} classified this galaxy as E$_{\rm pec}$, because
of the dust lane.  \citet*{hms56} and \citet{s61} classified it as an
S0$_3$. Based on photographic surface photometry, \citet{bcr82} show that the
object consists of bulge, lens, and exponential disc components,
characteristic of a lenticular galaxy. This easily rules out a classification
as (dwarf) elliptical, as several authors have proposed.

For many years the distance to NGC~404 was basically an unknown quantity.
Estimates ranged from 1.5\,Mpc \citep{bw76} to 10\,Mpc \citep{wh90}. It is
only recently that a firm distance to this object has been established.
\citet{tdb} used the surface brightness fluctuation method to come up with a
distance of 3.3 Mpc. This was confirmed independently by \citet{kara02} who
used the tip of the red giant branch and determined a distance of $3.15 \pm
0.32$\,Mpc.

Although NGC~404 is listed as a member of the loose group LGG~011
\citep{g93}, it is unlikely to be physically associated with the rest of the
group as at 3.3~Mpc \citep{tdb} the nearest member, NGC~598 (M\,33), is
located at 2.6 Mpc. Rather, NGC~404 is likely the nearest isolated S0 galaxy,
especially as there are no other known galaxies within 1.1~Mpc of it
\citep{kara02}. It is interesting to note that the field is only two degrees
away from the centre of the Piscis cluster of galaxies ($\rm \alpha =
01^h\,10^m \;\delta = 33.4^\circ$). A summary of optical data on NGC~404 is
given in Table~\ref{optical-datos}.

\section{Observations}

\label{obser}

NGC\,404 was observed with the NRAO Very Large Array (VLA)\footnote{The
National Radio Astronomy Observatory (NRAO) is operated by Associated
Universities, Inc., under contract with the National Science Foundation
(NSF).} on 1996 April 1 (C--array) and 1996 July 16 (D--array) in the 21 cm
line of neutral hydrogen (\ion{H}{1}). In C--array (with baselines between
100 and 3000 m, or 0.5 to 15 k$\lambda$) the object was observed for 132
minutes; in D--array (with baselines between 20 and 1000 m or 0.1 to 5
k$\lambda$) it was observed for 54 minutes. The absolute flux calibration was
determined by observing 3C\,48, which was assigned a flux density of 15.99
Jy. This calibrator was also used to derive the complex bandpass
correction. A nearby calibrator (0116+319) was used as secondary amplitude
and phase calibrator, and its flux was determined to be $2.635\pm0.005$ Jy.
Details about the observations can be found in Table~\ref{vla-datos}.  The
data were edited and calibrated separately for each array with the {\sc AIPS}
package\footnote{The Astronomical Image Processing System ({\sc AIPS}) has
been developed by the NRAO} using standard reduction techniques.

The systemic velocity of NGC~404 is $\sim -54\rm \,km\,s^{-1}$ at a Galactic
latitude of $-27.01^\circ$, hence some of the channels were confused with
Galactic emission (in particular between $-52$ and $-46 \rm \, km\,
s^{-1}$). Since Galactic emission shows its presence on predominantly large
scales, we were able to remove the most prominent emission features by
blanking the {\em uv} data out to $0.5 \,\rm k\lambda$ in these channels
(this affects only D--array).

The {\em uv} data were examined and bad data points caused by either
interference or cross talk between antennae were removed, after which
the data were calibrated. We Fourier--transformed our C-- and
D--configuration observations separately to assess their
quality. Finally, we combined both data cubes to form a single dataset
which was used for further analysis. The original bandwidth of the
data was 1.56\,MHz; for the rest of this paper we use a 0.69\,MHz wide
bandwidth around the spectral region of interest. The data have a
velocity resolution, after on--line Hanning smoothing, of 12.2~kHz
($\rm 2.57\,km\,s^{-1}$).

Channels with velocities higher than $\rm 16\,km\,s^{-1}$ and lower than $\rm
-128\,km\,s^{-1}$ were determined to be line--free. A total of 20 channels on
either side of the main profile containing no line emission were averaged to
form a ``dirty" continuum image, again avoiding the edge channels of the
bandpass. The average continuum level was subtracted from the {\em uv} data.

\section{\ion{H}{1} spectral channel maps}

\label{hi-line-chan-maps}

High ($\sim 15''$) resolution cubes were obtained using the task {\sc imagr}
with the parameter {\em robust}=--1 (where {\em robust}=--5 corresponds to
uniform weighting and {\em robust}= 5 to natural weighting).  With this value
we obtain the best compromise between the smallest beam (almost identical to
`uniform') and the lowest noise (almost identical to `natural'). Cleaning was
applied down to a flux level of 2 times the average rms noise in the maps and
in an area large enough to include all the prominent emission.
Fig.~\ref{canalesgrises} shows the resulting channel maps (before
blanking). Notice that most traces of contamination by Galactic emission have
been successfully removed. To save space, we display the average of two
consecutive channels in each panel over the velocity range where line
emission is present. The heliocentric radial velocities (in $\rm km\,s^{-1}$)
are plotted in the upper right--hand corner of each panel. The beam size
($15.2''\times14.4''$) is indicated in the top left--hand panel, left bottom
corner (the beam size is so small that it cannot be properly reproduced in
our maps). Already at first glance we can see quite strong emission (for an
S0 galaxy).  The resulting high resolution cubes
($512\times512$ pixels $\times 57$ channels at $15.2''\times14.4''$) were
smoothed to a spatial resolution of $25'' \times 25''$ (HPBW) in order to
bring out the faint, extended structures (referred to henceforth as low
resolution cubes). Table~\ref{vla-result} lists the characteristics of the
resulting data cubes.

\section{Global \ion{H}{1} content}

\label{contenidogas}

\begin{figure}
\plotfiddle{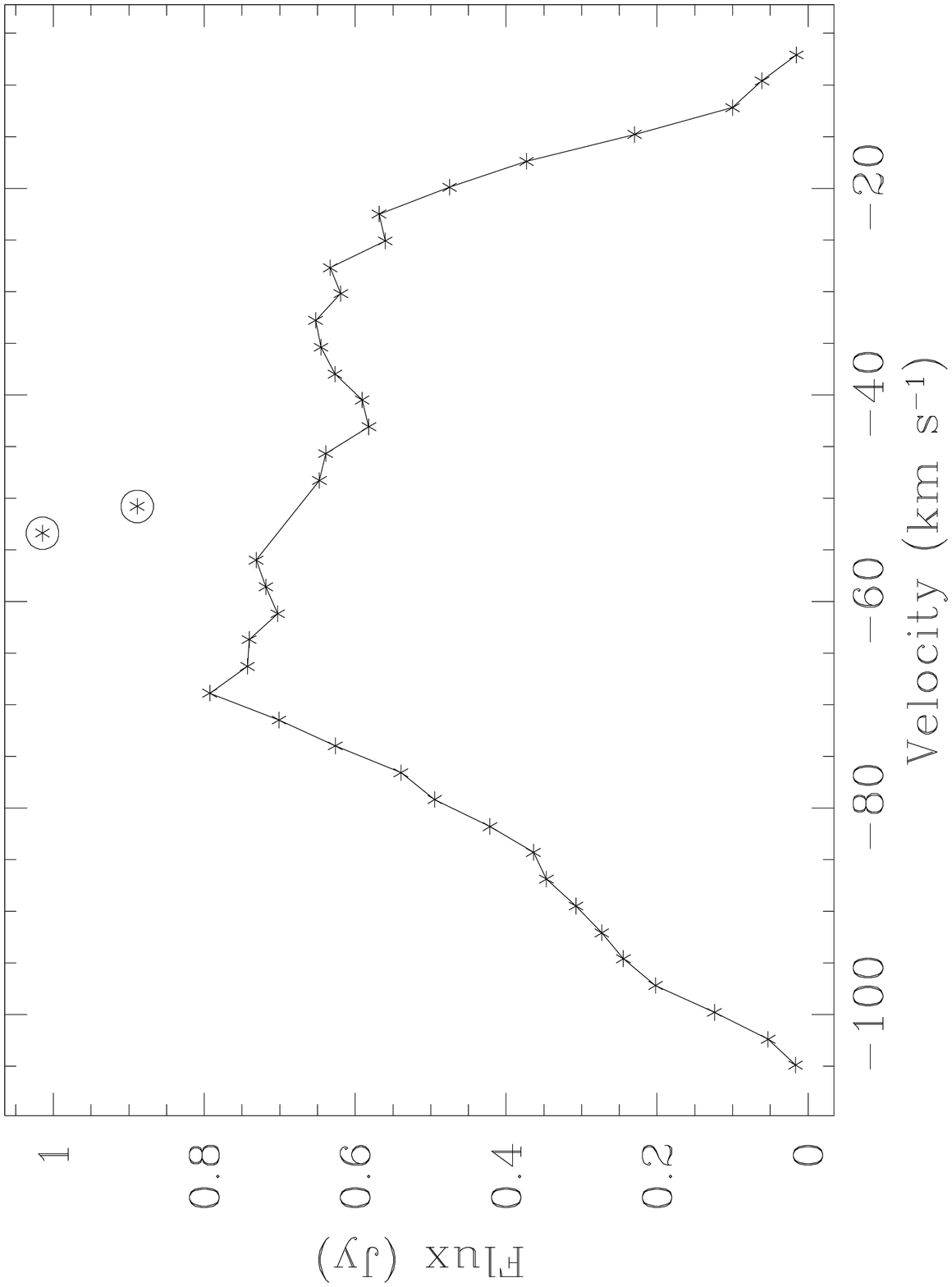}{7.5cm}{270}{45}{45}{-200}{250}
\caption{Global \ion{H}{1} profile. The two encircled points represent values
  which are contaminated by Galactic emission that cannot be
  completely removed from the data cube.}
\label{cuernos}
\end{figure}

The global \ion{H}{1} profile was obtained by integration over the
individual high resolution channel maps (note that this can be safely
done as all continuum emission was subtracted in the {\em uv}
plane). Before the summation, the flux at each position was corrected
for attenuation by the primary beam of the array. The resulting global
profile is shown in Fig.~\ref{cuernos}.  From the global profile we
derive a total \ion{H}{1} flux of $\rm 59.3\pm 1.7 Jy\,km\,s^{-1}$, a
systemic velocity of $\rm -53.9\,km\,s^{-1}$ and a global \ion{H}{1}
velocity profile width at 20\% and 50\% of the maximum of 85 and 65
$\rm km\,s^{-1}$ (not corrected for inclination, instrumental
resolution, turbulent or $z$ direction gas motion; the error is of
order one channel width or $2.5\,km\,s^{-1}$).

In order to derive the correct \ion{H}{1} flux in each of the channel maps,
one has to keep in mind that the determination of \ion{H}{1} fluxes of
extended emission in interferometric maps is not straightforward. As
explained by \citet{jm95}, fully cleaned maps do not exist, and any cleaned
map consists of the sum of two maps: one containing the restored clean
components and the other, the residual map. In the former, the unit is Jansky
per clean beam area, and in the latter Jansky per dirty beam area. Following
\citet{wb99}, we used our cleaned data cube of NGC~404 in order to calculate
the real flux of our channels. We calculate this real flux ({\em G}) using
{\em G=(D$\times$C)/(D$-$R)}, where {\em C} is the cleaned flux, and {\em R}
and {\em D} are the ``erroneous" residual flux and the ``erroneous" dirty
flux over the same area of the channel map, respectively. For a full
discussion on this topic, the reader is referred to the appendix in
\citet{jm95}. We followed this procedure for a selected sample of channels,
and compared the results with our estimates for the real flux which were
obtained by a direct, deep clean. The agreement is to within a few percent,
indicating that using our cleaned maps we get values which are very close to
the real flux.

Our integrated \ion{H}{1} flux is in fair agreement with that estimated by
\citet{bw76} using the Effelsberg 100--m telescope who arrived at a value of
$\rm 42\,Jy\,km\,s^{-1}$. It should be noted, however, that the latter value
was reported as being highly uncertain due to contamination of the single
dish spectra by foreground emission from the Galaxy and could be off by up to
25\%.

Using~:
\begin{equation}
\mhi\, [M_\odot] = 2.356\times10^5 \times D^2\, [{\rm Mpc}] \times \int
F_i\,{\rm d}v\,\rm [Jy\,km\,s^{-1}]
\end{equation}
we find a total \ion{H}{1} mass of $\mhi=1.52\times10^8 \, M_\odot$, which is
an order of magnitude less than the mean mass found by \citet[and references
therein]{vdvw91} for their sample of S0 galaxies. The mean \ion{H}{1} mass
for 14 S0 galaxies is $\sim 2.0\times10^9 M_\odot$, using H$_0 = 75\rm \,km\,
s^{-1}\, Mpc^{-1}$. The \ion{H}{1} mass--to--blue luminosity ratio, $\mhi/L_B
= 0.22$, which is independent of distance, on the other hand is high for an
S0 galaxy, and much closer to that found for later type spirals \citep{rh94}.

A quick calculation shows that NGC~404 is underluminous for an S0 galaxy, as
indicated by the Tully--Fisher (1977) relation for this object. Correcting
the global profile measured at the 20\% level for instrumental broadening and
for turbulent and $z$ direction motions by $-12.6\,\rm km\,s^{-1}$ (following
\citealt{rh84}) and assuming an inclination of $11^\circ$ (the value that we
have found for the inner part of the disc) we estimate a peak to peak
velocity for gas in the disc of $380\,\rm km\,s^{-1}$ or a circular velocity
of $190\,\rm km\,s^{-1}$. The required $B$ luminosity in order to have
NGC~404 fall on the Tully--Fisher curve is more than an order of magnitude
higher (2.5 to 3 magnitudes) than the value we assumed here and would imply
a distance of order 20 Mpc which is completely ruled out by the distance
determination of 3.3 Mpc by \citet{tdb} and \citet{kara02}.
 
\section{\ion{H}{1} distribution and Velocity field}

After Fourier transforming the {\em uv}--data the resulting data cube was
conditionally blanked, accepting as genuine all emission at levels above $2
\sigma$ rms noise in the data cube smoothed to a resolution of $25''$, and
requiring that the emission is seen in at least 3 consecutive channels.  Maps
of the \ion{H}{1} column density distribution and velocity field were
obtained by taking the zeroth and first moment of the spectra at each pixel
along the velocity axis.  This was done in both the high and low resolution
cubes.

\subsection{\ion{H}{1} distribution}

\begin{figure}
\epsscale{0.8}
\plotone{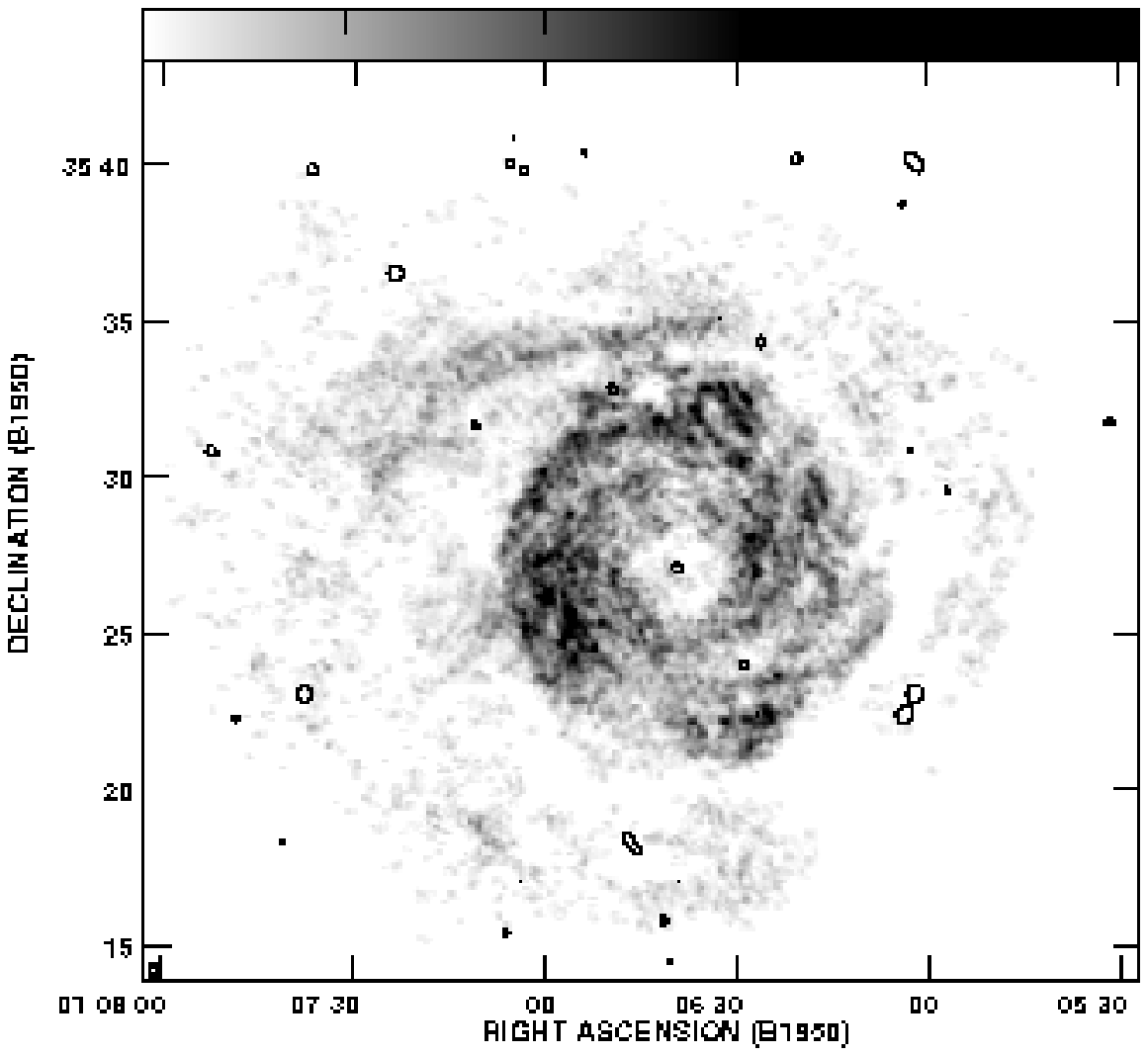}
\caption{Integrated neutral hydrogen map of NGC~404. Intensities range from 0
to 300 K\,km\,s$^{-1}$ or $5.5 \times 10^{20}$\,at\,cm$^{-2}$. Continuum
sources over $5\sigma$ (0.75 mJy) are superimposed on the emission line map.}
\label{integrado}
\end{figure}

The \ion{H}{1} is distributed as a bright doughnut with an inner diameter of
$\simeq 100''$ and an outer diameter of $\simeq 400''$ (see
Fig.~\ref{integrado}), that is, from R$\sim R_{25}$ to R$\sim 4R_{25}$.  At
considerably lower signal--to--noise level an outer ring can be discerned,
out to twice that diameter.  The ellipticity of that ring is higher than that
of the disc, suggesting a different inclination.  Our VLA \ion{H}{1} data
imply a mass of $\mhi \simeq 1.14 \times 10^8 M_\odot$ in the ``galaxy''
(R$<400''$) and $\mhi \simeq 0.38 \times 10^8 M_\odot$ in the ``annulus''
(R$>400''$), that is, roughly a distribution of 3/4 of the total mass in the
galaxy and 1/4 in the outer ring.  There is considerable fine--scale
structure visible, but no large scale pattern such as spiral arms.  The
doughnut is very nearly circular, suggesting an almost face--on
orientation. This is corroborated by isophotal fits to the optical
image. \citet{bcr82} derive values for the inclination and position angle as
a function of radius, from R$\sim 2''$ to R$\sim 88''$. The values they found
vary from $158^\circ$ to $75^\circ$ in PA and from $0.238$ to $0.062$ in
ellipticity. \citet{wh90} use a constant value of $i=16^\circ$ and a position
angle, PA $= 100^\circ$.

\begin{figure}
\epsscale{1}
\plotfiddle{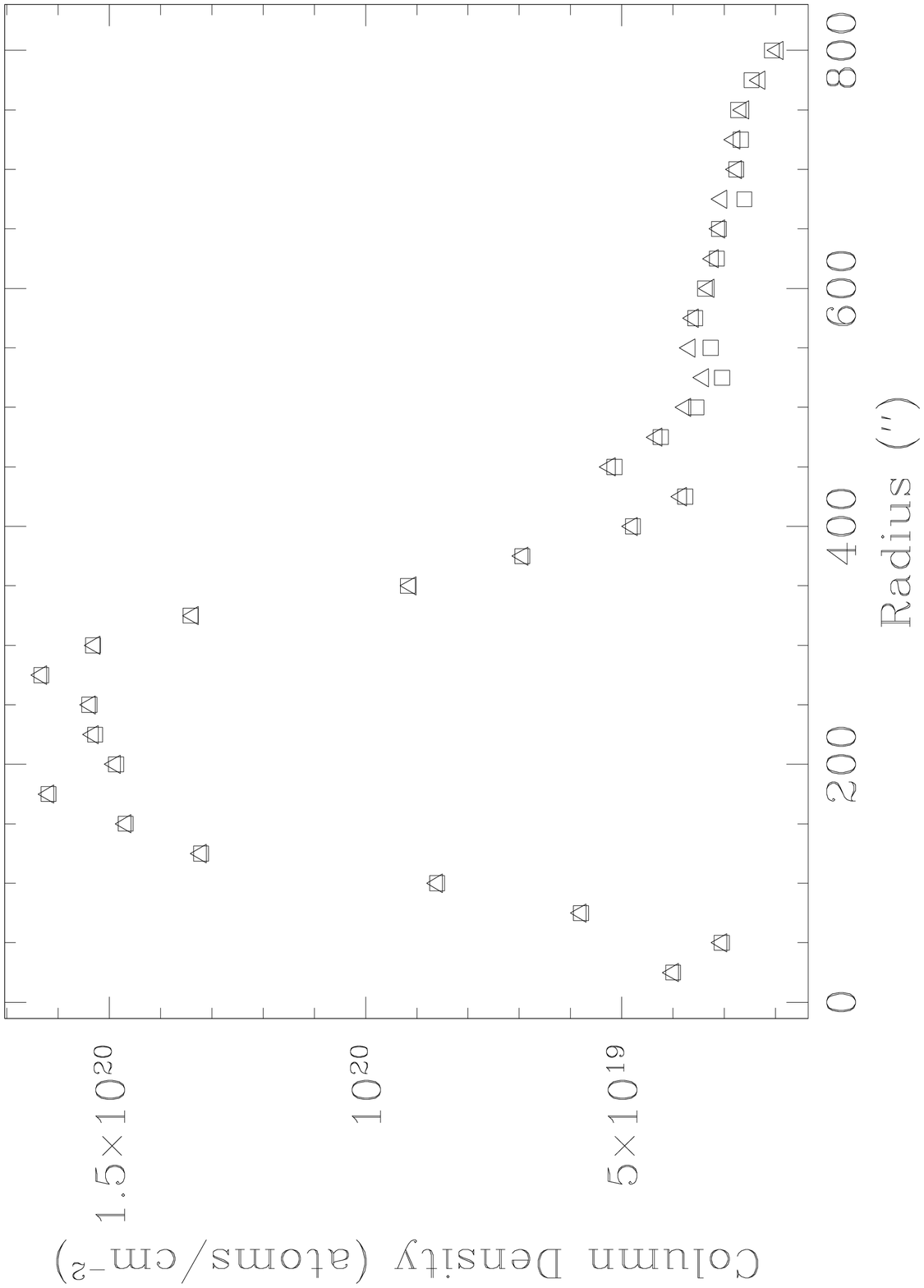}{8cm}{270}{45}{45}{-200}{250}
\caption{Mean column density as a function of radius, obtained by averaging
  in rings. In this graph one can clearly appreciate the doughnut.
  ($\triangle$ correspond to fits to the galaxy assuming a flat rotation
  curve; $\Box$ assume a Keplerian decline)}
\label{fluxdensfit}
\end{figure}

The azimuthally averaged radial distribution of the \ion{H}{1} surface
density, corrected to face--on, $\sigma_{\mhi}$, is shown in
Fig.~\ref{fluxdensfit} for the full resolution data. The profile was derived
from the \ion{H}{1} column distribution map by averaging in circular rings in
the plane of the galaxy, centered at the position of the nuclear continuum
source, with the task {\sc iring} of GIPSY\footnote{The Groningen Image
Processing System (GIPSY) is distributed by the Kapteyn Astronomical
Institute, Groningen, The Netherlands}. The values for PA and $i$ are those
derived based on the velocity field for a flat and a Keplerian rotation curve
(see section \ref{curvarot} for details). There are no significant
differences, though, between the two sets of inclination angles in
determining the \ion{H}{1} surface density.

The surface--brightness vs. radius profile cannot be adjusted to any of the
distribution functions that usually characterize these profiles (gaussian or
modified Hubble profile). Neither corresponds to a truncated disc either, the
decrease beyond R=$400''$ being much steeper than an exponential.

\subsection{Velocity field}

\begin{figure}
\plotfiddle{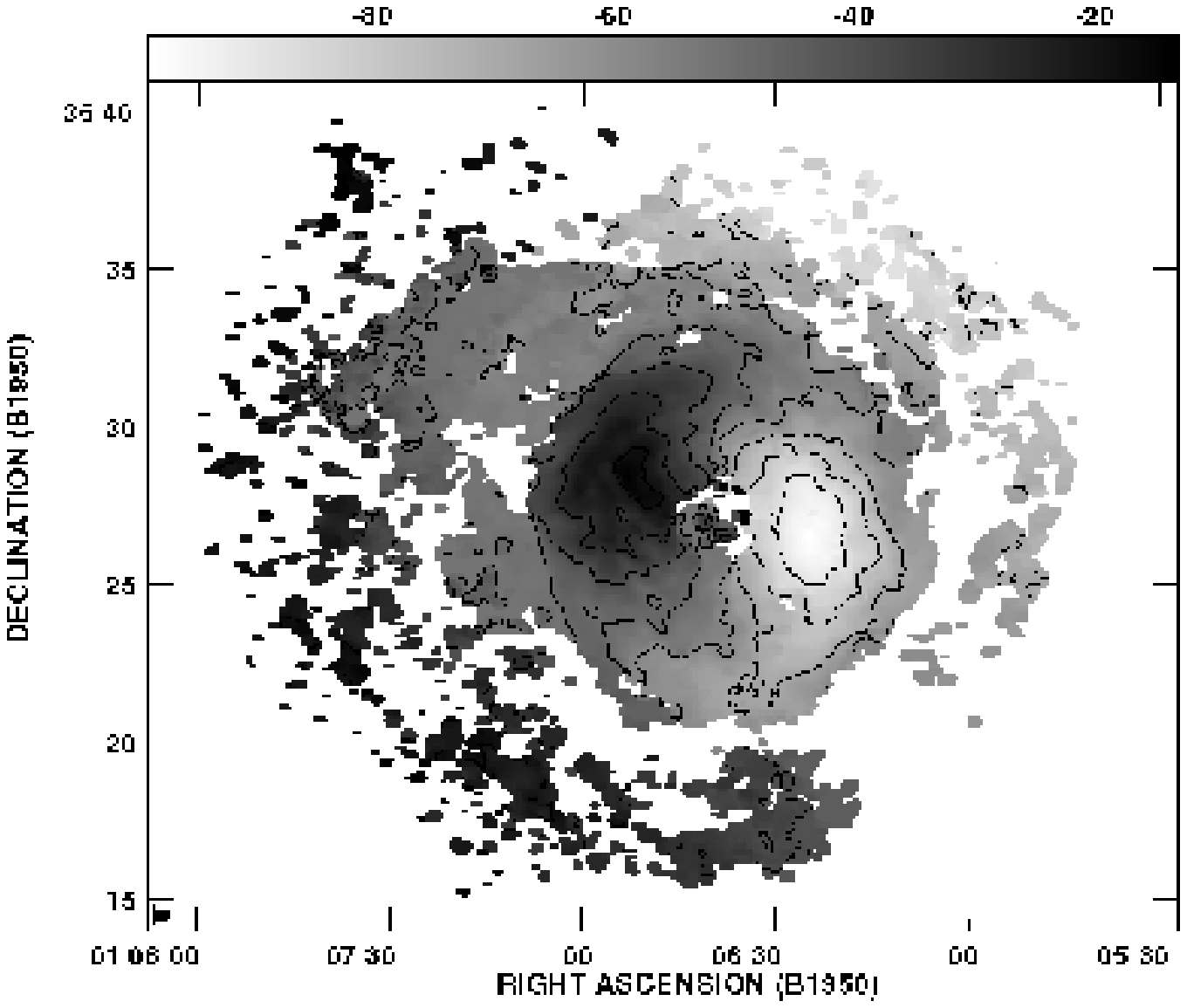}{5.0cm}{0}{40}{40}{-250}{-80}
\plotfiddle{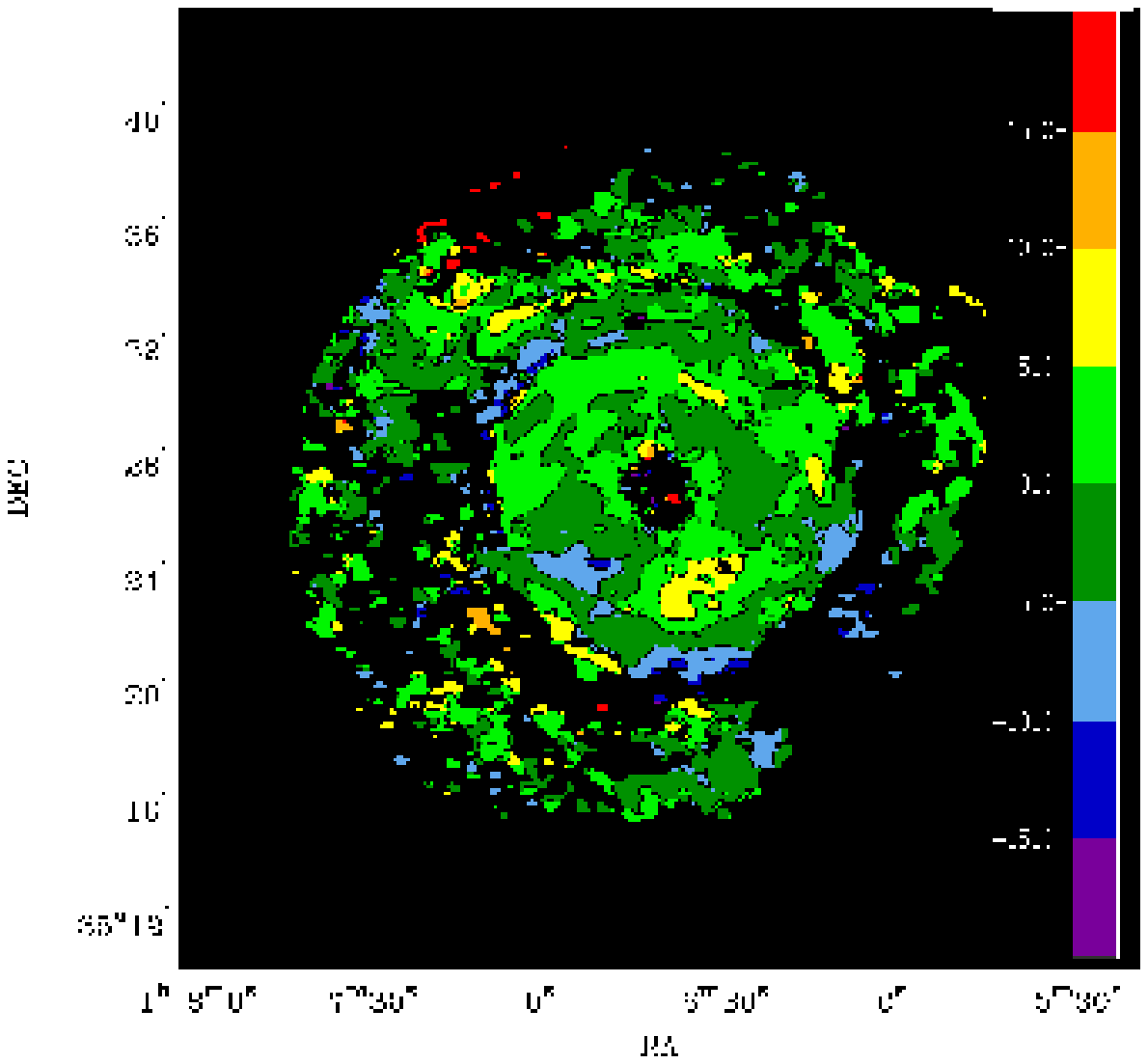}{0cm}{0}{40}{40}{-30}{0}
\caption{Left: Velocity map. The observed radial velocities range from $-100$
  to $-20$ km\,s$^{-1}$, the western half being the receding side. Right:
  Residual map obtained after subtracting a Keplerian model from the data.
  Contours range from~: $-10$ to $10$ $\rm km\, s^{-1}$ in steps of 5\,$\rm km\, s^{-1}$}
\label{mapa-vel}
\end{figure}

\begin{figure}
\plotfiddle{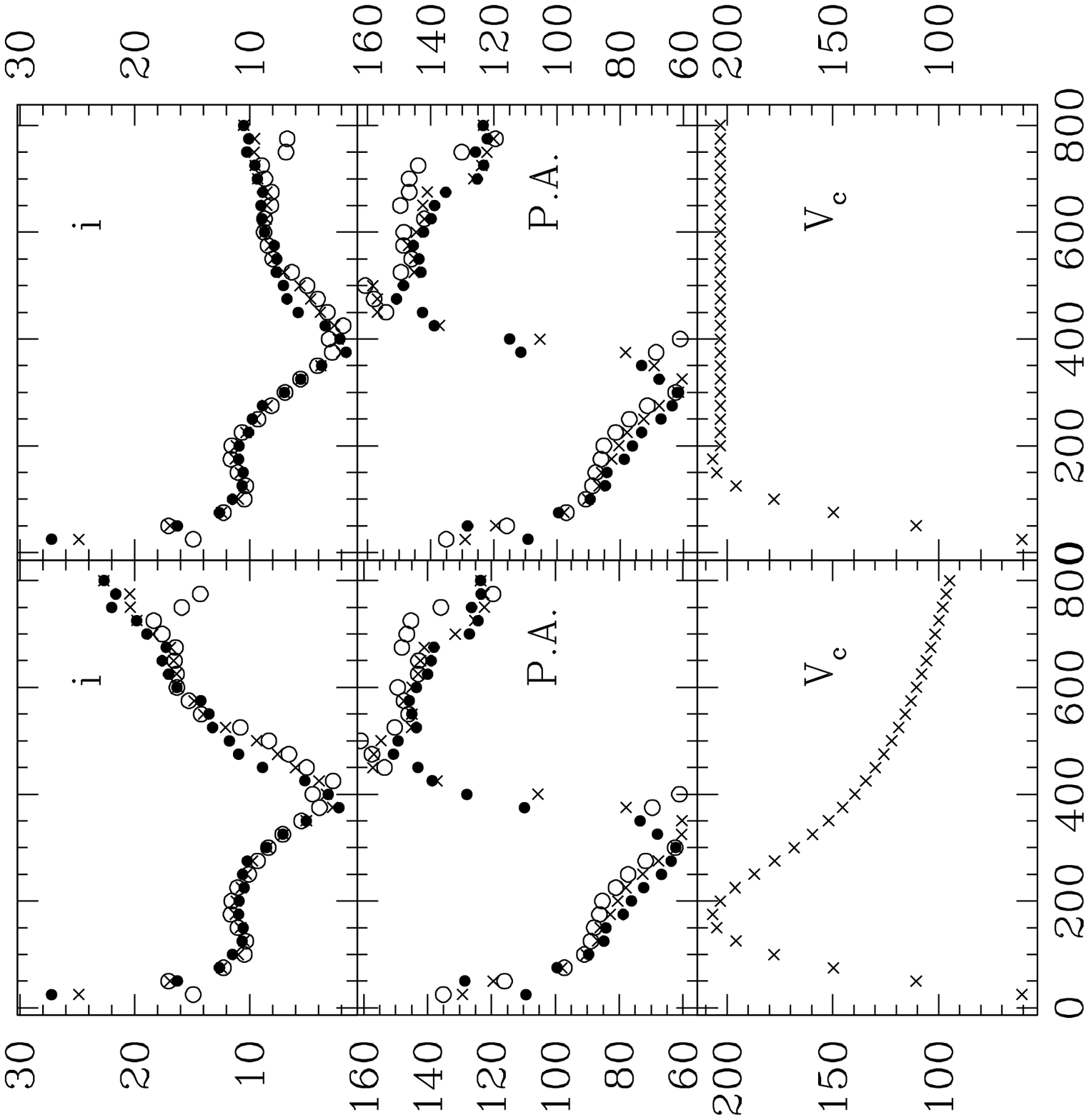}{13cm}{270}{80}{80}{-250}{450}
\caption{Tilted--ring model fits to the velocity field of NGC\,404 showing
solutions for a Keplerian (left) and Flat (right) rotation curve.  The axes
are labeled as follows: P.A. and $i$ are in degrees, V$_c$ is in
km\,s$^{-1}$, and the radial distance from the centre is in arcsec. 
Crosses represent the result of the average for both sides. Open circles correspond
to the fit of the receding and filled circles to that of the approaching side, respectively.} 
\label{three-fits}
\end{figure}

A velocity field, based on the smoothed, $25''\times25''$ channel maps, is
presented in Fig.~\ref{mapa-vel}. We note that the isovelocity contours are
closed, and that the velocities of the external annulus do not form a smooth
continuation with those of the main body. Also, the line of nodes varies as a
function of radius implying the presence of a warp.  The projected radial
velocities range from $\sim -100$ to $\sim -20 \,\rm km\,s^{-1}$. The
velocities of the outer ring fall within the same range as those of the main
body, but the position angle shows a jump (see Fig.~\ref{three-fits}).

\section{Rotation curve}

\label{curvarot}

The velocity field in the inner region ($< 400''$) is sufficiently regular to
permit the determination of a rotation curve for which we used a tilted--ring
model as implemented by the task {\sc rotcur} in the GIPSY package
\citep*{rlw74,b89}.

The model assumes circular symmetry in each of a set of concentric, but not
necessarily coplanar rings centered on the nucleus.  The projected circular
velocity $V(r,\theta,i)$ (which is based on the observations) is related to
the real velocity $V_c(r)$, assuming circular orbits of the gas via:
\begin{equation}
V(r,\theta,i)=V_{\rm sys} + V_c(r) \sin(i) \cos(\theta - \rm PA) 
\label{vel-rot}
\end{equation}
where $V_{\rm sys}$ is the systemic velocity, $i$ is the inclination angle,
PA is the position angle of the receding major axis (the line of nodes) and
$\theta - \rm PA$ is the angle measured with respect to the receding major
axis of the object.

To find a set of these parameters that best represent the observed velocity
field {\sc rotcur} performs non--linear least squares fits to the observed
radial velocities in a set of concentric elliptical annuli in the plane of
the sky, each one beamwidth ($25''$ after smoothing, Fig.~\ref{mapa-vel})
wide along the major axis. Points on either side of the major axis are
included up to $\theta = 60^\circ$ (i.e. we excluded a sector around the
minor axis). Using best guesses for the systemic velocity ($V_{\rm sys}$) and
the kinematical centre of the object, fitted values for $i$, PA and an
approximate rotation curve were calculated. Once fits for $i$ and PA as a
function of radius were found, these values were held fixed and $V_{\rm sys}$
and the central position were fitted. We repeated this procedure a couple of
times until consistent values for the central position and $V_{\rm sys}$ were
arrived at. The systemic velocity of $-55.4\rm\, km\,s^{-1}$ and the
kinematical centre, $\alpha_{\rm 1950}= 01^h\, 06^m\, 39.40^s$ and
$\delta_{\rm 1950}= 35^\circ\, 27'\, 5''$, are found to be in good agreement
with published values. E.g., based on observations of $\rm
^{12}CO(1\rightarrow 0)$, \citet{s90} found $V_{\rm sys} = -58\,\rm
km\,s^{-1}$; \citet{wh90} derived $V_{\rm sys} = -56\,\rm km\,s^{-1}$.

We then proceeded to minimize the dispersion of $V_c(r), i$ and PA for each
ring. It should be noted that the solution for $V_c(r)$ and $i$ are coupled
through the factor $V_c(r)\cdot \sin(i)$ in Eq.~\ref{vel-rot}, and that this
coupling is very strong for low inclinations. In the case of NGC\,404 we
decided to adopt an inclination of $i = 11^\circ$ near a radius of $200''$,
close to the location of the peak of the observed rotation curve. This
inclination is based on an intensity weighted average of the ellipticity as a
function of radius of the \ion{H}{1} doughnut. This inclination, being
derived under the assumption that the \ion{H}{1} is distributed in a thin
disc, is likely a better approximaton of the true value than the inclination
quoted by \cite{bcr82}. With this choice of inclination, the observed peak
velocity translates to an intrinsic peak rotational velocity of about $200\rm
\,km\,s^{-1}$ at a radius of $200''$\footnote{ An inclination of $i =
11^\circ$ is a lower limit. A more relaxed value would be $i = 16^\circ$, as
used by \cite{wh89} which results in a peak velocity corrected for
inclination a factor of 1.4 lower, at $140\rm \,km\,s^{-1}$ and dynamical
masses a factor of 2 lower.}.

Not being able to independently fit $V_c(r)$ and $i$, we instead
considered two extreme solutions. the first one being a Keplerian
decline (model K). This model is inspired by the observed radial
velocity as a function of radius which follows in its outer parts an
r$^{-1/2}$ curve. Fits to the receding and approaching side of an
r$^{-1/2}$ decline agree to within $1\sigma$. We followed
\citet{ssy71} and constructed a Keplerian 
rotation curve, based on the sum of an inhomogeneous spheroid and a Toomre
disc with $n_T = 3$, scaled to match the peak velocity of our data. 

The second solution we adopted was for the rotation curve to remain
flat (model F) after reaching a peak rotational velocity of $200\rm
\,km\,s^{-1}$ at a radius of $200''$. 

\begin{figure}
\plottwo{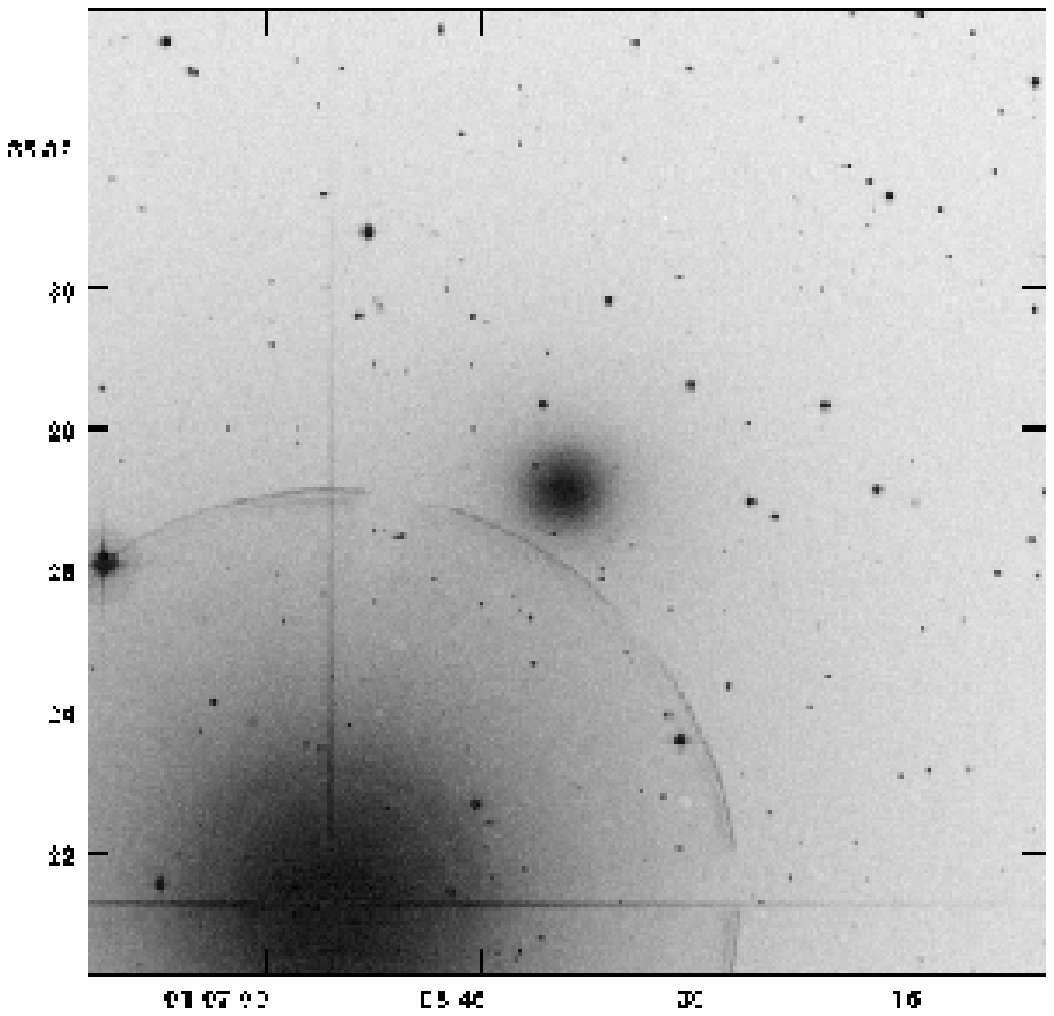}{delrio.fig08b.ps}
\caption{Display of an optical image taken from the DSS (left) and
  at the same scale, the \ion{H}{1} distribution (right). The bright star
  object to the south-east of NGC 404 in the DSS image is
  $\beta$And. The plus (+) indicates the center of NGC~404, the star
  ($\star$) represents the location of $\beta$And.
}
\label{both}
\end{figure}

\begin{figure}
\epsscale{0.8}
\plotone{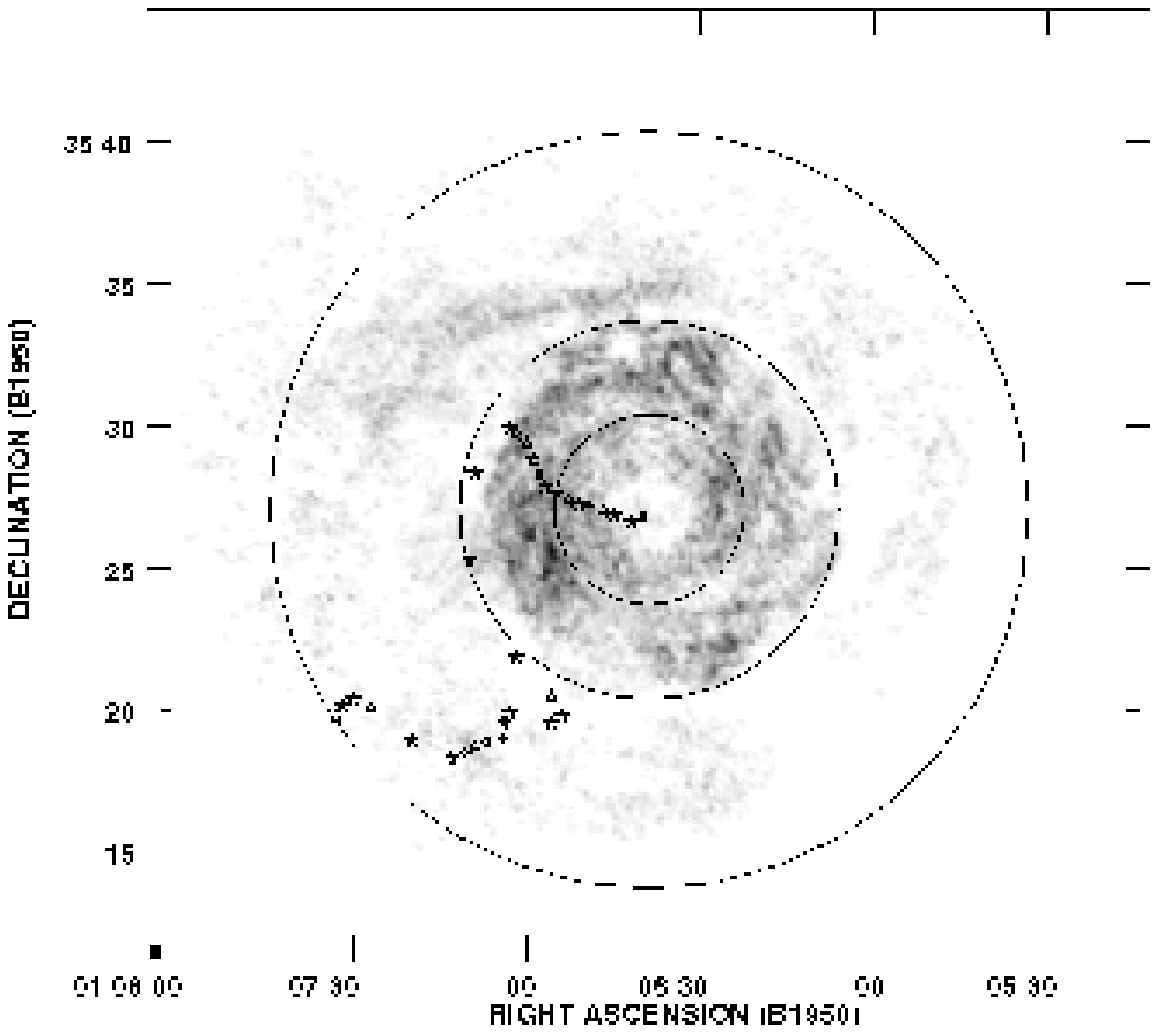}
\caption{The direction  of the P.A. has been drawn as five pointed stars. The circles
  represent, from the innermost to the outermost: R=$200''$, where the
  rotation curve begins to decline ($\approx 2$R$_{25}$), R=$400''$, which
  marks the outer rim of the doughnut, and R=$800''$, the last point of our
  velocity model. It can be seen that the P.A. is almost constant within
  $R_{25}$ ($\sim 200''$), at which radius the warp sets in.}
\label{punteli}
\end{figure}

With the rotational velocity thus fixed, we reran {\sc rotcur} and for
each model we left only as free parameters to be fitted the PA and
$i$. The results of these restricted fits are shown in
Fig.~\ref{three-fits}. In this figure the values for the receding and
approaching sides, as well as the average  over both sides is plotted. The
first thing we note is that for both models the position angles share
practically the same values, indicating that this parameter is insensitive to
either choice of rotation curve and probably reflects the true run of
position angle with radius. The two first points (R$\leq 50''$) are not
significant as they fall within the \ion{H}{1} hole (see Fig.~\ref{both}),
where there is scant gas and the fit can take almost any value. From R$>
50''$ to R$\sim 350''$ the value for the PA declines monotonically and almost
linearly from $100^\circ$ to $60^\circ$. This is depicted in an alternative
manner in Fig.~\ref{punteli}. There is a discontinuity at R$\sim 400''$
(close to $4R_{25}$). It is exactly around this radius that we reach the edge
of the main body of the galaxy and that we see the transition to the
annulus. Finally, the annulus shows the same tendency, but less pronounced by
a factor of 2, the PA decreasing from $160^\circ$ to $120^\circ$.

\begin{figure}
\plotfiddle{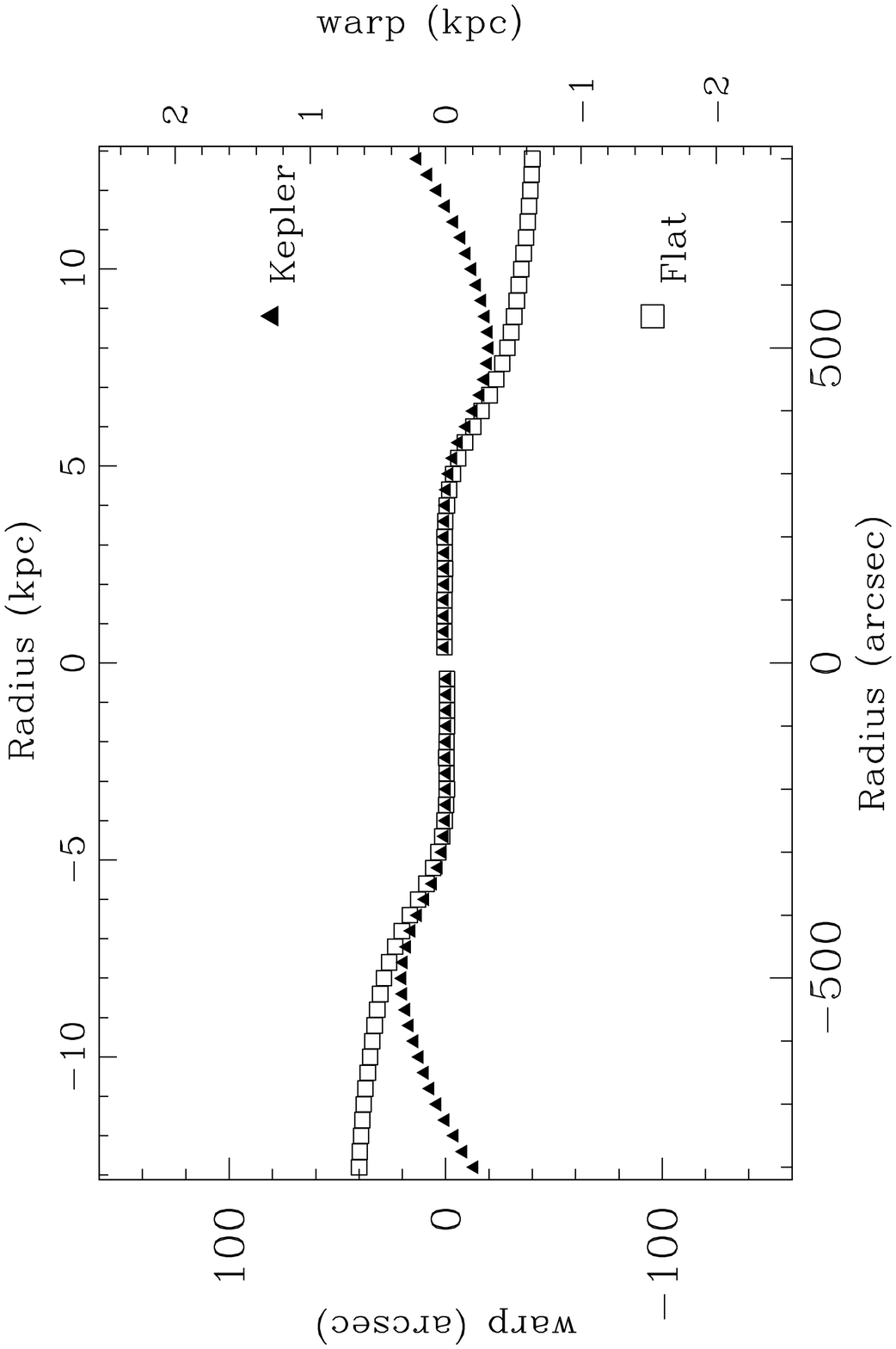}{10cm}{-90}{60}{60}{-225}{350}
\caption{Fig. a) A schematic view of both warps (for a Flat and Keplerian rotation
curves) as seen from an edge-on perspective, in the plane of the galaxy. The
$y$-axis has been enlarged to better visualize the differences between both
warps.}
\label{2warp}
\end{figure}

The behaviour of PA being so similar in both models, the most significant
differences then have to be a function of $i$, and mainly for R$>
200''$, where the functional behaviour of both rotation curves starts to
diverge. This is indeed what we find. Within the errors, the inclination for
the doughnut is constant, differences as small as $1^\circ$ being
sufficient to mask the difference between the K and F models. Beyond R$\sim
500''$ the inclination in the case of the K model has to increase whereas in
the case of the F model it remains roughly constant. Fig.~\ref{2warp} is an
attempt to visualize the shape and extent of the implied warp, both for a
Keplerian decline as well as for a Flat rotation curve.

\section{The Mass of NGC~404}

We can now proceed to derive the mass within the last measured point for both
the K and F model fits to the velocity field. In the case of a Keplerian
decline and assuming for simplicity a spherical mass model: $M_{\rm T} =
2.326 \times 10^5\,V_{\rm rot}^2 \,R$, ($M_{\rm T}$ in $M_\odot$, $V_{\rm
rot}$ in $\rm km\,s^{-1}$ and $R$ in kpc), we derive a total mass $M_{\rm T}
\approx 3 \times 10^{10} M_\odot$. This is assuming the peak rotational
velocity of $200 \rm \,km\,s^{-1}$ is reached at $200'' (\sim 2R_{25})$ after
which point the rotation curve falls off. In that case the global $M/L_{B}$
ratio is $\sim 44 \, M_\odot/L_\odot$. This is quite a high value and a
factor of 1.5 higher than the already substantial values encountered by
\citet{vdvw91}.

If we now consider a flat rotation curve, taking as the last measured point a
value of $\sim 800''$, the mass turns out to be four times higher, $\sim 12
\times 10^{10} M_\odot$, and the mass--luminosity ratio would reach a
staggering $\sim 175\, M_\odot/L_\odot$.

Although the $M/L_B$ values are high, NGC~404 cannot be considered a massive
galaxy. If we take the total mass enclosed within $R_{25}$ ($R=1.65 \,\rm
kpc$ and $V=160 \rm\, km\,s^{-1}$) we find $M_{\rm T}=0.98\times 10^{10}\,
M_\odot$, less than 10\% of the mean mass encountered within the same radius
in the galaxies making up the sample of \citet{vdvw91}. In that case the
mass--to--light ratio is $14.4 \, M_\odot/L_\odot$. Although this is still
high, it is more in line with that found in Early Type galaxies.

\section{Continuum 1.4 GHz}

To check for the possible presence of low level continuum emission, we
extracted 34 channels at either side of the combined {\em uv}--data, free of
line emission. The 68 channels were averaged and Fourier--transformed to the
image domain and cleaned in a similar manner as the \ion{H}{1} maps. Contours of
the 1.4 GHz map are shown superposed on the \ion{H}{1} map in
Fig.~\ref{integrado}.

\citet{bw76} found as upper limit for the continuum emission $S_c = 3$ mJy.
We detect radio continuum emission from the nucleus of NGC~404 at the
20$\sigma$ level. The source is unresolved and has a flux density of
$3.6\pm0.3$\,mJy.  We find a position of $\alpha_{1950} = 01^h\,06^m\,39.25^s
\pm \,0.03^s$ and $\delta_{1950} = 35^\circ\, 27'\, 07.0'' \pm\, 0.3''$. The
errors are formal errors to the fit only. The radio and optical positions
coincide to within the accuracy of the latter.

The radio power of the nuclear source (at an assumed distance of 3.3\,Mpc) is
given by $\rm \log P_{21} (W\,Hz^{-1})=17.9$. According to \citet{h80} and
\citet{hk82}, about 80\% of S0/a to Sab--type galaxies have a central
continuum source with a total power similar to, or greater than, $\rm \log
P_{21} (W\,Hz^{-1})= 19.4$. In the study of S0 galaxies by \citet[and
references therein]{vdvw91}, 90\% of the galaxies have a power higher than
those indicated by \citet{h80}, even the lowest value found for their weakest
object (log\,P$=18.75$ and assuming H$_0 = 75\rm \,km\,
s^{-1}\, Mpc^{-1}$) is more than 7 times higher
than the value we find for NGC\,404.  The luminosity of the nuclear source is
more than an order of magnitude lower than the Galactic supernova remnant
Cas~A and comparable to that of the Crab Nebula.

There are $\sim$ 50 point sources with $S_{21} > 3 \sigma$, but only one
(besides the nuclear source) has $S_{21} > 15 \sigma$. This source is located
at the outer limit of the bright emission of \ion{H}{1}, and thus falls
beyond the optic limit.

\citet{mea98} found a nuclear UV spectrum characteristic of a very young ($<
3$\,Myr) or at most moderately old ($\sim 5$\,Myr) burst of star formation.
They conclude that the continuum UV source of this LINER is a stellar
cluster.  Its low luminosity could be explained with 2--6 O stars, although
the real number probably is higher after correcting for extinction.  With our
results we can calculate the massive star formation rate (following
\citealt{c92}) as SFR($ M\ge 5\,M_\odot)\,\approx 2 \times 10^{-4}
M_\odot\,\rm yr^{-1}$, a very low value if compared with the Milky Way ($0.3
- 0.5 \,M_\odot \,\rm yr^{-1}$) or E+A galaxies ($0.5 \,M_\odot \,\rm
yr^{-1}$, \citealt{mo00}) or even when compared with other S0's, where the
massive star formation rate is of the order of $\sim 0.1\,M_\odot \,\rm
yr^{-1}$ (from data given by \citealt{vdvw91}).

\section{Origin of the \ion{H}{1}}

The question about the origin of the gas in galaxies of Hubble type S0 hasn't
been fully answered yet. For example, gas lost by evolving stars is likely to
be retained by these fairly massive systems, offering a possible explanation
for the origin of the gas which is encountered. But in that case, one would
expect a far larger proportion of S0s to be detected in \ion{H}{1}. In the
following we will look at the different scenarios to explain the origin (from
internal as well as external sources) of the detected gas in NGC~404.

\subsection{Burnt--out disc}

This scenario assumes that lenticular galaxies are former spirals, that
consumed efficiently their gas content in the process of star formation
\citep{lea80}. In that case, the gas--rich S0s could be lenticulars which,
although they consumed the gas in the inner region, somehow retained some gas
in the outskirts. This \ion{H}{1} is then the left--over primordial gas from
which the galaxy was formed. Due to the low density of the gas in the outer
regions, it never formed stars there.

Like many S0s, NGC~404 presents a central hole in its \ion{H}{1} distribution
giving it the appearance of a doughnut with an inner and outer radius of $\sim 
1-4 R_{25}$. This extent is larger than that found in other S0s. 
The azimuthally averaged neutral atomic gas surface
density, $\sigma_{\rm H{\mbox{\scriptsize\sc i}}}$, in this ring is
$\sigma_{\rm H{\mbox{\scriptsize\sc i}}} = 1.2\, M_\odot \rm pc^{-2}$, too
low by at least a factor of 5 for a efficient star formation to proceed, and of the 
same order as that encountered in many gas rich S0's. We used Kennicutt's criterion
\citep{ken89} to look in a bit more detail in to the this issue. Using the observed one 
dimensional velocity dispersion of the gas and observed rotation curve, we find that 
the gas surface density is too low by an order of magnitude for gravitational collapse
of the gas to occur.

This implies that this scenario runs into trouble on various fronts. Firstly, the neutral
gas surface density is way too low for star formation to take place. In fact, one would
expect not only that a close to critical surface density would be found, but also
that the peak gas density is reached at the edge of the stellar counterpart. In fact,
we find it farther out, near $1.5R_{25}$, i.e., far from the optical edge of
the galaxy which is where the purported SF zone ends. Also, within this 
framework other observed features of the gas, such
as the change in position and inclination angles, or the presence and
morphology of the outer annulus are difficult to account for. Hence, we
consider it improbable that the \ion{H}{1} gas in NGC~404 is mainly
left--over primordial gas.

\subsection{Mass loss of evolved stars}

It is well established that the ISM is constantly replenished through
contributions of evolved stars which inject gas and dust into the ISM.  This
process is also occurring in S0s and indeed, the amount of gas isn't
negligible: ${\rm d}M/{\rm d}t \sim 0.015(L_V/10^9 L_\odot)[M_\odot\,\rm
yr^{-1}]$ \citep{fg76}. This implies that the majority of S0s should be gas
rich and that this is the rule rather than the exception. As we find the
contrary, there must exist a mechanism which rids S0s from this gas. Star
formation doesn't seem a viable candidate as little SF activity is recorded
in S0s. Similarly, a galactic wind seems to be ruled out, as is interaction
with, e.g., an intracluster medium, NGC~404 finding itself
isolated. Alternatively, the gas which is lost inherits the velocity
dispersion of the stars leading to temperatures which are in the X--ray
regime.

For a typical S0 with $\mhi/L_B \lesssim 0.1\, M_\odot/L_\odot$ and a blue
luminosity, $B$, of order $10^9$ to $10^{10}\,L_\odot$ we obtain a reasonable
estimate for the time needed (a few times $10^9$ yr) to collect the observed
amount of gas as a result of mass loss from evolved stars. However, in the
case of NGC~404, its luminosity is very low ($\sim 7\times\,10^8\,
L_\odot$). This, combined with an \ion{H}{1} mass to blue luminosity ratio
which is higher than the majority of S0s studied thusfar ($0.22 \,
M_\odot/L_\odot$), results in an estimated time scale of order a Hubble
time. So despite the fact that theoretically the ISM detected in NGC\,404
could be accounted for by steady mass loss since it formed, there are serious
shortcomings to this potential explanation. Mass loss by evolved stars
wouldn't answer the question why most of the ISM is found beyond $1.5R_{25}$ either,
where star formation is and has been negligible. Moreover, it would be
difficult to explain where it got its angular momentum from.

\subsection{External Origin}

The former scenarios offer unsatisfactory explanations for the origin of
\ion{H}{1} in NGC~404. They especially fail to explain the fact that the
outer ring is tilted. In fact, this is one of the stronger arguments for
postulating an external origin of the gas.

The \ion{H}{1} content could be the result of the merger of a dwarf irregular galaxy
with NGC~404. If we assume $\mhi/L_B\, \sim 0.6$ as a value representative
for normal dwarf galaxies ($0.3\leq \mhi/L_B \leq 1.0$; \citealt{he99}) we
find that a dwarf with $M_B \sim -15.5$ has a large enough gas supply to
account for the observed amount of $1.52\times10^8 M_\odot$ of
\ion{H}{1} gas.

Assuming $M_{\rm T}/L_B \sim 4\,M_\odot/L_\odot$ \citep{he99}, the total mass
of the dwarf, in a single merger, would have been $M_{\rm T}\sim 10^9
\,M_\odot$. The total mass of NGC~404 is, at least, $\sim
3\times10^{10}\,M_\odot$. The mass of the dwarf then would have been only
$\sim 3\%$ of that of NGC~404, and the merger would have had no major
dynamical effect on the latter.

In section \ref{curvarot} we assumed that the gas could be described as
moving in concentric annuli within the potential well of the galaxy. If left
to their own design, such annuli tend to settle in a preferred plane, the
plane of the galaxy. Following \citet{tsc82} we have calculated the time
scale ($\tau_d$) for settling to a preferred plane of two annuli in NGC~404,
one at R=$200''$ and one at R=$800''$, assuming that the donor galaxy left
its gas in a plane inclined by about $40^\circ$ with respect to NGC~404. For
R=$800'', \, \tau_d \approx 3.5\times10^9$ yr, whereas for R=$200'', \,
\tau_d \approx 0.9\times10^9$ yr (which corresponds to $\sim 63$ rotational
periods). As can be appreciated in Figs.~\ref{three-fits} (the plots of $i$
versus radius) and \ref{2warp}, the gas in the ring at R=$200''$ has already
settled in the plane of the galaxy. Now if we assume that the gas has been
accreted during just one interaction, this must have happened at least
$0.9\times10^9$ yr ago. This time lapse agrees fairly well with that
necessary for the ring at R=$800''$ to settle from an original inclination of
$i\sim 40^\circ$ to its current value of $i\sim 11^\circ$ (assuming Keplerian
rotation). These arguments restrict considerably the time elapsed since the
interaction.  If it were much smaller than $0.9\times10^9$ yr, the gas of the
inner ring (the doughnut) wouldn't have had time to settle into the plane of
the galaxy whereas if it were much larger, the outer ring would have settled
within the plane of the disc of NGC\,404.

If we assume that the gas currently encountered in NGC~404 was transferred in
a fly--by encounter, the culprit must have moved outside of a volume with a
radius of 1.1 Mpc as there are no objects within 1.1~Mpc \citep{kara02}. This
would imply a relative velocity for the possible intruder of order $1200
\rm\, km\,s^{-1}$, an uncomfortably high value. This rather argues in favor
of a catastrophic merger.

In case the rotation curve is flat (model F), the dynamical time scale,
$\tau_d$, is slightly shorter, $\sim 0.5\times10^9$ yr for R$=200''$ and
comes out to be $\sim 2\times10^9$ yr for R$=800''$.

\section{Discussion}

Numerical simulations of hierarchical galaxy formation within the $\Lambda$
Cold Dark Matter ($\Lambda$CDM) paradigm and carried out with sufficient
resolution (in mass) to distinguish dwarf galaxies (\citealt{klea99,
mooea99}) predict the existence of numerous low-mass halos. Observations,
however, come up with numbers much smaller than predicted. One of the
possible explanations which has been invoked is that these halos are too
small to produce star formation, or that their luminosity is so low that this
makes them difficult to detect. In those systems where star formation has
taken place, $M/L_B$ ratios tend to be high \citep{tullyea02}. These systems
are thought to appear in groups. In contrast, NGC~404 is an isolated
galaxy. \citet{kara02} suggest that this galaxy may represent the final stage
of consecutive merging of members of a former group of galaxies. In this
case, it could be the dominant member of a group of small galaxies, some of
which, rich in neutral gas, have contributed the material which extends up to
almost $8R_{25}$.

The high $M/L$ ratio confirms that this is an extreme case where matter
(bright and dark) is highly concentrated. This high mass concentration also
fits in with the picture of the orbits of gas captured through a merger event
decaying within a few Gyr and aligning with the plane of the galaxy
\citep{cvg91}. This implicitly assumes that there is no extended dark halo to
maintain gas orbits in a warp for a long period of time.

Regarding the apparently falling rotation curve, there are several other
galaxies (\citealt{cp90} (NGC~7793), \citealt{cvg91} (NGC~2683, NGC~3521),
\citealt{ps95} (a few galaxies among their 967 spiral galaxy sample),
\citealt{o96} (NGC~4244), \citealt{jbh96} (NGC~4138), \citealt{hs97}
(11 galaxies), \citealt{sth99} (the Milky Way, among others), and
\citealt{cgnb02} (NGC~7631, KUG~2318+078)) in which the rotation curve
begins to decline close to $2R_{25}$. However, in most of these
cases the \ion{H}{1} measurements don't go much further than this
radius and this might be the reason why no convincingly declining
rotation curves have been found to date. 

So, Keplerian declines are rare. \cite{hs97} list 11 galaxies for which
with some confidence one can state that the rotation curve falls
significantly (by 50--100\,km\,s$^{-1}$). The most extreme case, and
in fact a system quite similar to NGC~404, is the galaxy NGC~4138 which
was observed by \cite{jbh96}. \ion{H}{1} was detected out to 2.5
optical disc radii (16 scale--lengths). This gas is counterrotating
with respect to the main stellar body, suggesting recent capture of a
gas rich satellite, like in the case of NGC~404. But unfortunately,
also in this case difficulties in fitting the rotation curve prevents
the authors from unambiguously confirming a Keplerian decline.  

In the case of NGC~404 Nature is playing tricks on us and
decided to put this object close to face--on, making it impossible to solve
separately for the inclination and rotation speed in the plane of the
galaxy. Hence, we decided to analyse two extreme options for the
intrinsic rotation curve: a flat rotation curve out to the last measured
point (model F) and a curve following a Keplerian decline (model K).

If the rotation curve remains flat, then NGC~404 is one of the ``darkest
galaxies'' known with an M/L of 175 and a fraction of 92\% of its mass being
due to Dark Matter. It would also be one of very few galaxies for which a
rotation curve out to $\sim 8R_{25}$ (48 scale-lengths in the $B$--band) has
been determined.  The only other galaxy for which data up to $\sim 8R_{25}$
(15 scale-lengths) are available is DDO~154 \citep{cb89}.  On the other hand,
if this galaxy really follows a Keplerian decline, it would be
unique. In that case, the extent of the DM halo 
would reach out to 3.2 kpc ($\sim 2R_{25}$). One argument which
would favour Keplerian decline is that the warp geometry makes more sense and
qualitatively fits the picture of accreted material settling within the
gravitational potential of the galaxy, the outermost inclined annuli taking
longer than the more inward located ones. An additional argument is
that for model K the M/L ratio is more reasonable.

Finally, the preference for a Keplerian decline is reinforced by the fact
that in the case of a flat rotation curve, the inclination of the outer
annulus is the same as that of the inner part of the galaxy. This is at odds
with the shape of the outer annulus which clearly gives the impression of
having a larger ellipticity which instead would imply a higher
inclination. This is exactly what we find when fitting a the Keplerian
rotation curve. In this case we find that 68\% of the mass in NGC~404
is in the form of Dark Matter. 

\section{Conclusions}

NGC~404 is an exceptionally gas rich early--type galaxy.  The \ion{H}{1} is
distributed in two separate systems, a doughnut which extends from $R_{25}$
or $100''$ to $4R_{25}$ (1.6 to 6.4 kpc for a distance of 3.3\,Mpc) and an
annulus of much lower surface brightness which extends from $4R_{25}$ to
roughly $8R_{25}$.

We derive a total \ion{H}{1} mass of $1.52\pm 0.04\times10^8\,M_\odot$ of
which 75\% resides in the doughnut and the rest in the outer annulus. This
translates to a distance independent $\mhi/L_{B}$ ratio of 0.22 in solar
units which is high for an S0 galaxy knowing that the median for S0's
galaxies is $0.04 \,M_\odot/L_\odot$ whereas for Sb galaxies this is $\sim
0.22\,M_\odot/L_\odot$ \citep{rh94}.

The velocity field shows a steeply declining rotation curve. Unfortunately,
the galaxy is oriented almost face--on and, assuming circular rotation of the
gas in annuli, good fits can be obtained when forcing both a flat rotation
curve {\em and} a rotation curve that falls off like a Keplerian. In both
cases the position angle (PA) of the kinematical major axis varies with
radius by $\sim 40^\circ$.  In the Keplerian decline model, the gas in the
outer annulus is warped, the annulus moving to a less face--on orientation
with radius.

A total mass of $3\times10^{10}\, M_\odot$ is found when using the Keplerian
rotation curve. This implies a global $M_{\rm T}/L_{B}$ ratio of $\sim 44$ in
solar units, which is exceptionally high. This is partly due to the fact that
the blue luminosity of the galaxy is $\sim 15$ times lower than average for
an S0. For a flat rotation curve, a lower mass limit of $12\times10^{10}\,
M_\odot$ or four times higher is derived, leading to a consequently four
times higher total mass--luminosity ratio of $\sim 175\, M_\odot/L_\odot$.

The nuclear source is detected at 20\,cm. If nonthermal, its luminosity of
$\rm \log P_{21} (W\,Hz^{-1})=17.9$ is a factor of ten lower than Cas~A and
more in line with that of the Crab nebula. If dominated by star formation,
the SFR is a low SFR($ M\ge 5\,M_\odot)\,\approx 2 \times 10^{-4}
M_\odot\,\rm yr^{-1}$, 2--6 O stars being enough to explain the observed
luminosity.

Given the mass and luminosity of NGC~404, we are dealing with a dwarf
lenticular galaxy. Assuming Keplerian decline, the DM halo is compact and
ends near $2R_{25}$. Our preferred explanation for the origin of the ISM in
this system is that it has been accreted through a catastrophic merger with a
gas--rich, dwarf irregular system, less than a tenth of the mass of NGC~404
which has taken place anywhere between $0.5-1.0\times10^9$ yr ago (no other
objects having been encountered within a volume of 1.1\,Mpc radius). Gas at
small radius has already settled within the plane whereas gas in the outer
annulus is still gravitating towards the midplane of the disc defined by the
S0, resulting in a warp of the gas distribution.

Although none of our conclusions depends strongly on the rotation
curve being flat or declining, it is of course extremely frustrating
that it hasn't proven possible to decide between these two extreme
models, especially since the observed radial velocities decline so
strongly. This leaves NGC~404 as {\em potentially} the first galaxy
with a rotation curve that shows such a pronounced decline over such a
large extent which, if confirmed, would be of extreme importance.

\acknowledgments

We gratefully acknowledge a careful reading by an anonymous referee
which has improved the presentation of this paper. This project was
supported by Mexican CONACyT grants n$^\circ$ 33026--E to MSdR and
n$^\circ$ 27606--E to EB. This research has made use of the NASA/IPAC
Extragalactic Database (NED), which is operated by the Jet Propulsion
Laboratory, California Institute of Technology, under contract with
the National Aeronautics and Space Administration.

\clearpage

\begin{deluxetable}{ll}
\tablewidth{300.0pt}
\tablecaption{Optical parameters of NGC\,404}
\tablehead{}
\startdata
Morphological type & S0 \\
$\alpha_{1950}$ & $\rm 01^h 06^m 39.0^s$ \\
$\delta_{1950}$ & $+35^\circ 27' 6.0''$ \\
Isophotal major diameter D$_{25}$ & $3.5'$\tablenotemark{a} \\
Exponential scale length $B$ $\alpha^{-1}$ & $16.56''$\tablenotemark{b} \\
Exponential scale length $V$ $\alpha^{-1}$ & $129.5''$\tablenotemark{c} \\
Inclination $i$ & $11^\circ$\tablenotemark{d}\\
Position Angle P.A. & $100^\circ$\tablenotemark{d}\\
Adopted distance & 3.3 Mpc\tablenotemark{e} \\
Corrected total $B$ magnitude $\rm B^{0,i}_{T}$ & 10.93\tablenotemark{f} \\
Apparent blue magnitude $\rm m_{B}$ & 11.21\tablenotemark{f}\\
Total blue luminosity ($\rm M_\odot = 5.43$) & $6.8\times 10^8 \rm L_{B_\odot}$ \\
\tablenotetext{a}{NED} 
\tablenotetext{b}{del R\'{\i}o (in preparation)}
\tablenotetext{c}{\citet{bba98}}
\tablenotetext{d}{Values for $r<200''$}
\tablenotetext{e}{\citet{tdb}}
\tablenotetext{f}{RC3}
\enddata
\label{optical-datos}
\end{deluxetable}

\begin{deluxetable}{lll}
\tablewidth{360.0pt}
\tablecaption{VLA observing log}
\tablehead{}
\startdata
VLA configuration & C & D \\
Date of observation & 1-Apr-96 & 16-Jul-96 \\
Time on source & 132.5 minutes & 54 minutes \\
Total bandwidth & 1.56 MHz & 1.56 MHz \\
No. of Channels & 128 & 128 \\
Velocity resolution & 2.57 $\rm km\, s^{-1}$ & 2.57 $\rm km\, s^{-1}$ \\
Central velocity & $-57$ $\rm km\, s^{-1}$ & $-57$ $\rm km\, s^{-1}$  \\
\enddata
\label{vla-datos}
\end{deluxetable}

\begin{deluxetable}{ll}
\tablewidth{300.0pt}
\tablecaption{Characteristics of the datacubes}
\startdata
                    & High resolution \\
Half power beam     & $15.2''\times 14.4''$ \\
Noise per channel   & 1.1 mJy\,beam$^{-1}$ \\
Conversion factor (1mJy\,beam$^{-1}$) & 2.7 K \\
\ \\
                    & Low resolution \\
Half power beam     & $25''\times 25''$ \\
Noise per channel   & 0.65 mJy\,beam$^{-1}$ \\
Conversion factor (1mJy\,beam$^{-1}$) & 1.0 K \\
\enddata
\label{vla-result}
\end{deluxetable}

\clearpage

\onecolumn
\begin{figure}
\plotone{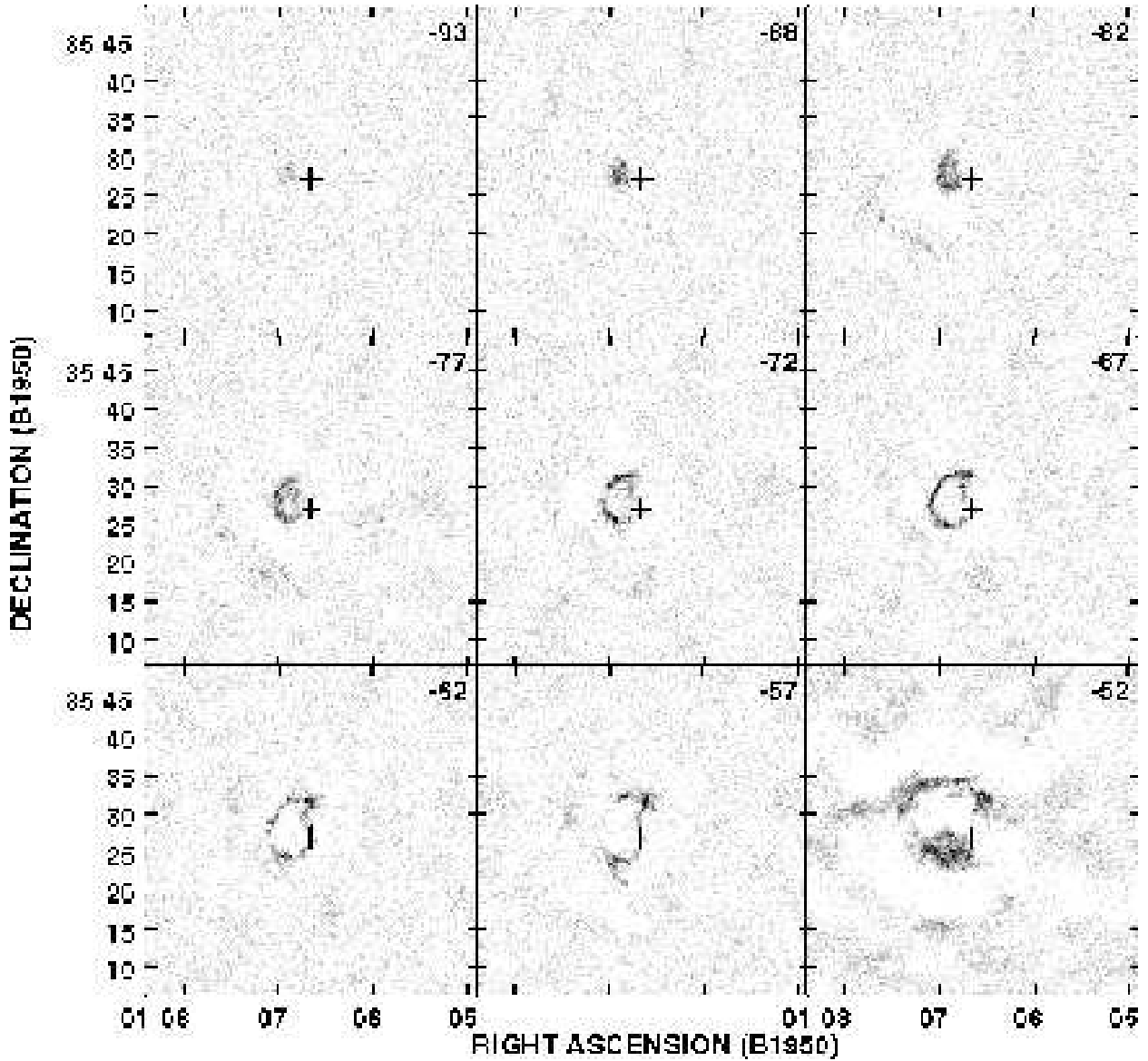}
\caption{\ion{H}{1} line channel maps of NGC\,404 after subtraction of the
continuum emission and cleaning. The spatial resolution (HPBW) is $15''.2
\times 14''.4$. In order to save space each panel is the average of two
neighboring channels. The heliocentric velocity of each channel has been
indicated in $\rm km\,s^{-1}$. The hatched circle in the upper left panel
shows the beam size. The grey scale runs from $\rm 0\, mJy\,km\,s^{-1}$ to
$5$ mJy$\rm \,km\,s^{-1}$. The cross indicates the optical and radio
continuum centre.} \addtocounter{figure}{-1}
\label{canalesgrises}
\end{figure}

\begin{figure}
\plotone{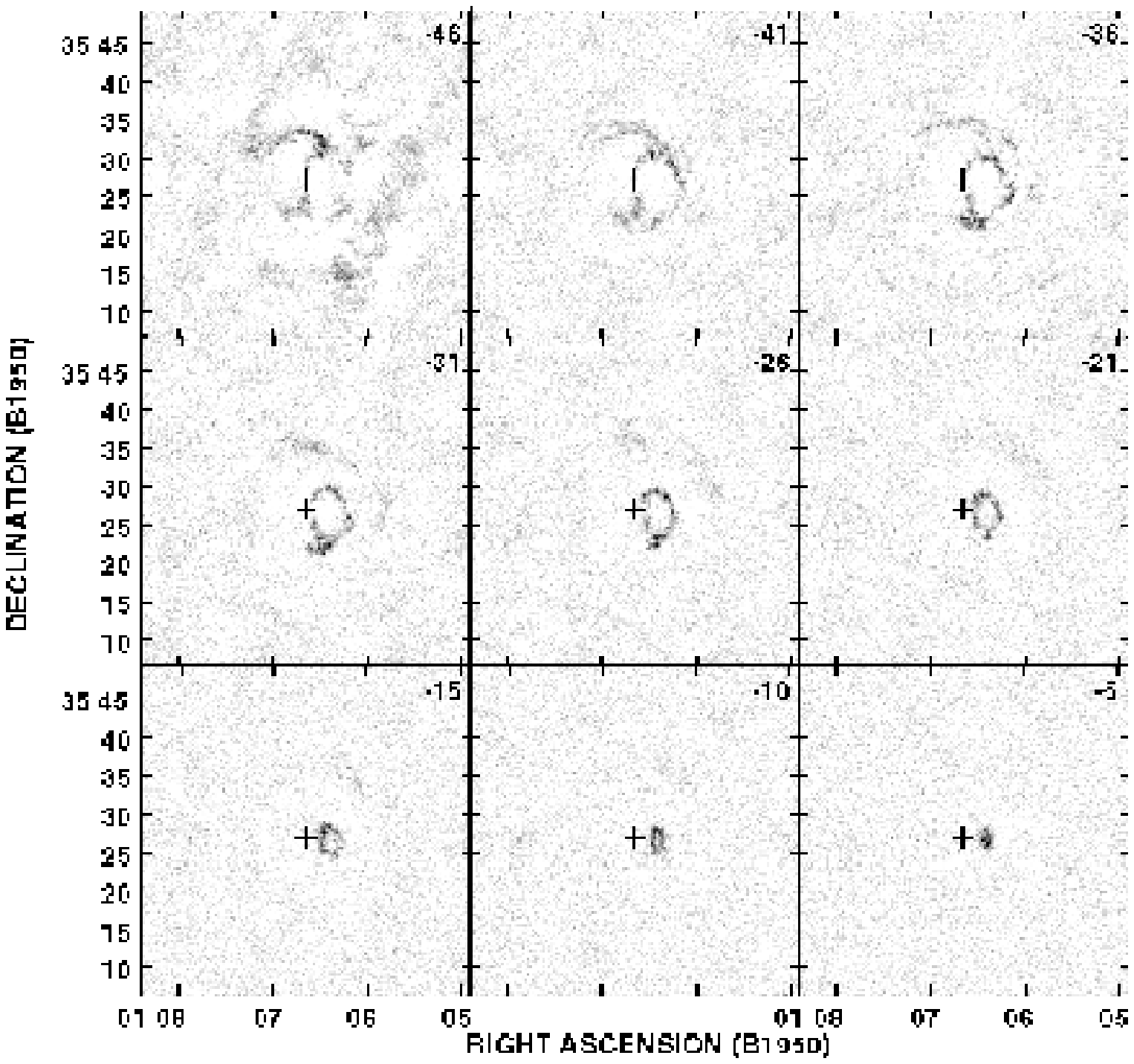}
\caption{{\em Continuation.-} Coordinates are with respect to B1950.0 equinox.}
\end{figure}

\epsscale{1}

\begin{figure}
\plottwo{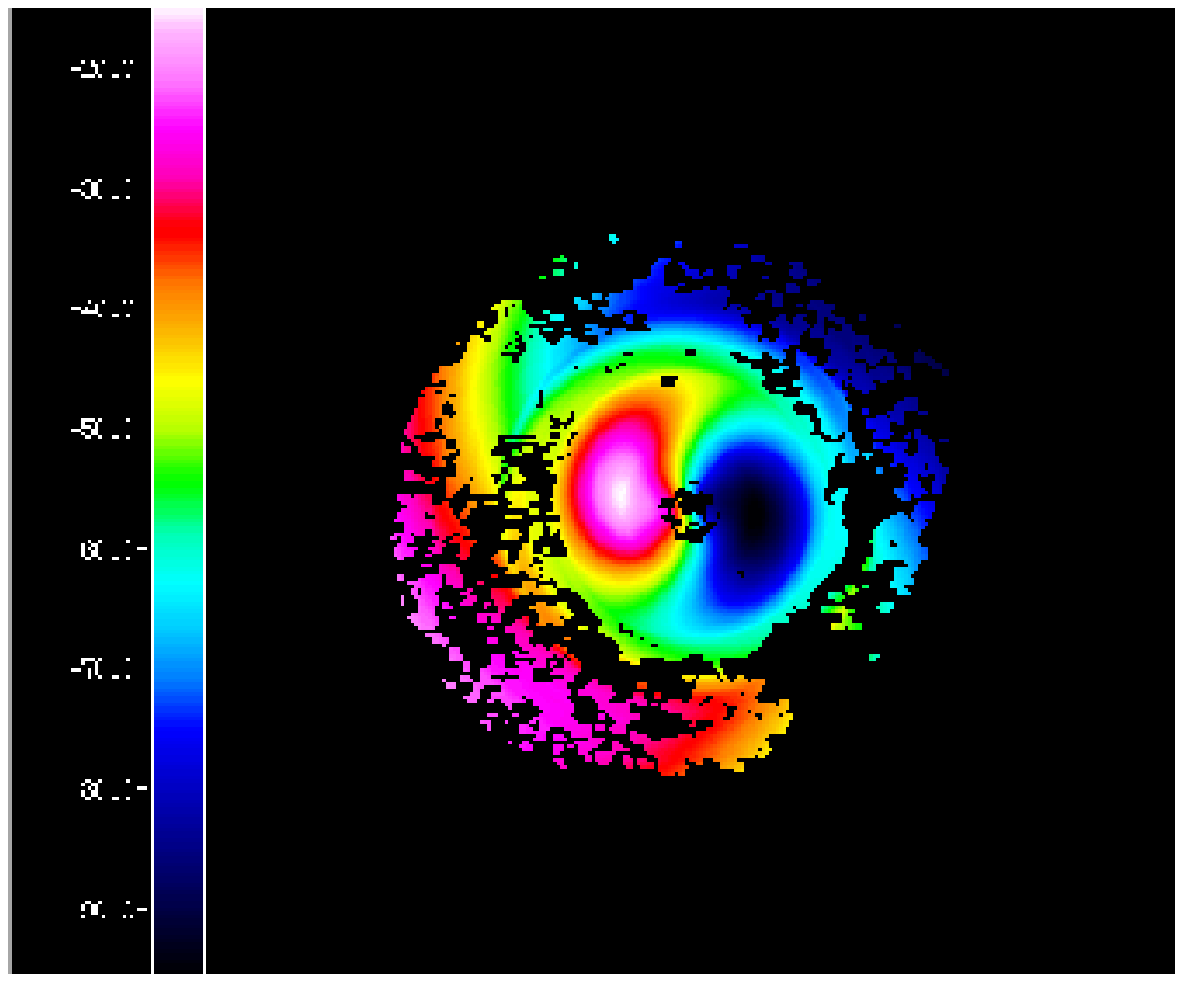}{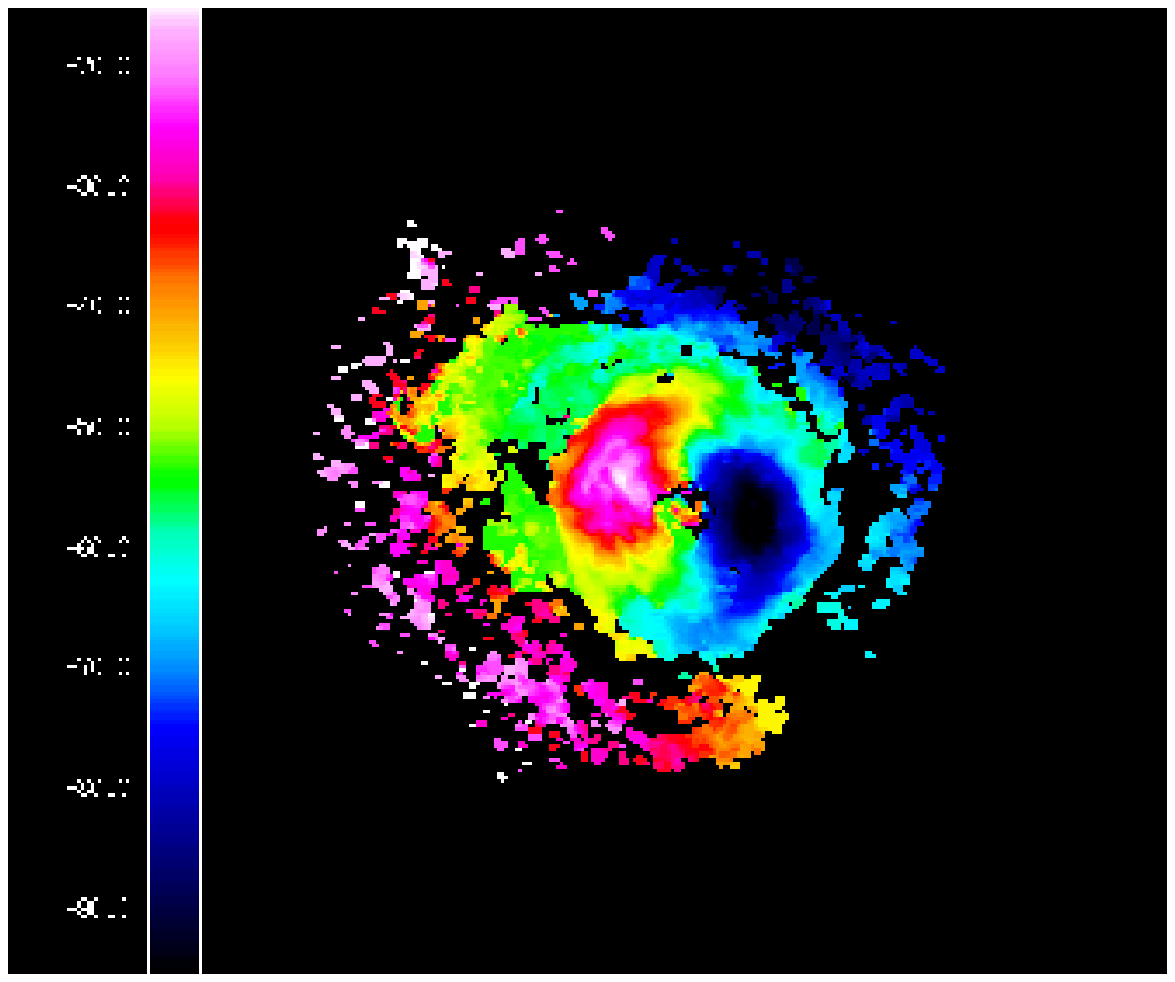}
\caption{Left: Model velocity field based on a rotation curve exhibiting a
  Keplerian decline, beyond a radius of 3.2 kpc. Right: Observed velocity
  field. The color scheme is the same for both panels.}
\label{model_velcol}
\end{figure}

\epsscale{1}

\epsscale{1}

\newpage

\epsscale{1}
\addtocounter{figure}{-1}
\begin{figure}
\plotfiddle{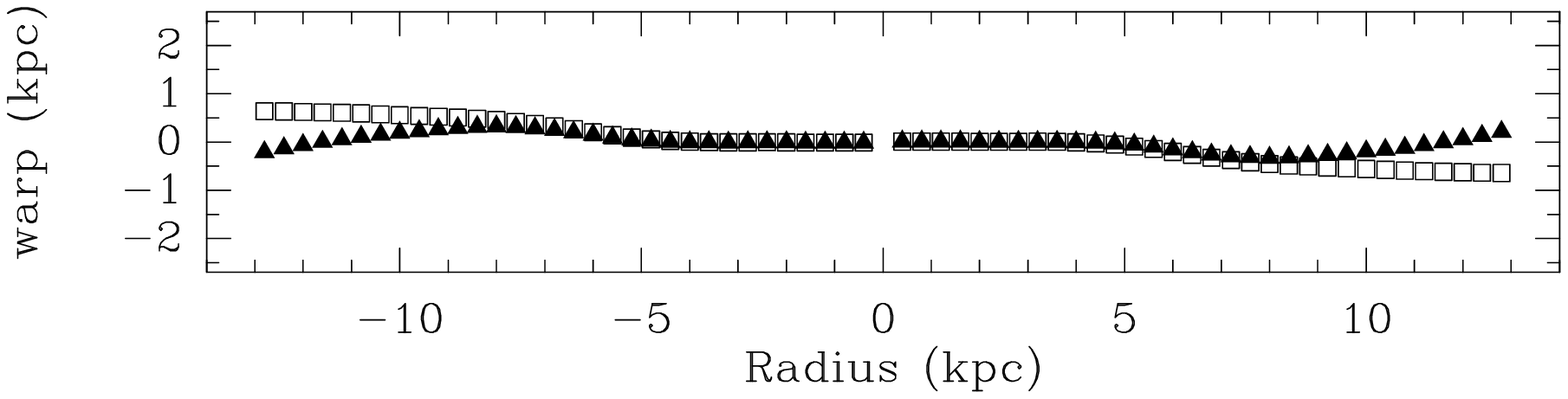}{5cm}{0}{90}{90}{-270}{-70}
\caption{Fig. b) As above, but with both
axes having the same scale to better visualize the amplitude of the warp
against the size (in \ion{H}{1}) of the galaxy.}
\label{2warpb}
\end{figure}

\end{document}